\documentclass[journal, transmag]{IEEEtran}
\ifCLASSINFOpdf
  \usepackage[pdftex]{graphicx}
  \DeclareGraphicsExtensions{.pdf,.jpeg,.png}
\fi
\usepackage{siunitx}
\usepackage{amsmath}
\usepackage[colorlinks=true,allcolors=blue]{hyperref}
\newcommand{\vect}[1]{\mathbf{#1}}
\newcommand*\diff{\mathop{}\!\mathrm{d}}

\makeatletter
    \@ifdefinable{\etal}{\def\etal/{\textit{et al.}}}
\makeatother

\begin{document}

\title{Transition Jitter in Heat Assisted Magnetic Recording by Micromagnetic Simulation}
\author{\IEEEauthorblockN{
Harald Oezelt\IEEEauthorrefmark{1},
Alexander Kovacs\IEEEauthorrefmark{1},
Johann Fischbacher\IEEEauthorrefmark{1},
Simon Bance\IEEEauthorrefmark{2},
Mark Gubbins\IEEEauthorrefmark{2} and
Thomas Schrefl\IEEEauthorrefmark{1}}
\IEEEauthorblockA{\IEEEauthorrefmark{1}Center for Integrated Sensor Systems, Danube University Krems, Wr. Neustadt 2700, Austria}
\IEEEauthorblockA{\IEEEauthorrefmark{2}Research \& Development, Seagate Technology, Derry BT48 0BF, United Kingdom}
\thanks{Corresponding author: H. Oezelt (email: harald.oezelt@donau-uni.ac.at).
\textcopyright 2017 IEEE. Personal use of this material is permitted. Permission from IEEE must be obtained for all other uses, in any current or future media, including reprinting/republishing this material for advertising or promotional purposes, creating new collective works, for resale or redistribution to servers or lists, or reuse of any copyrighted component of this work in other works. 
DOI: \href{http://dx.doi.org/10.1109/TMAG.2017.2709840}{10.1109/TMAG.2017.2709840}}
}

\IEEEtitleabstractindextext{
\begin{abstract}
In this paper we apply an extended Landau-Lifschitz equation, as introduced by Ba\v{n}as \etal/ for the simulation of heat-assisted magnetic recording. This equation has similarities with the Landau-Lifshitz-Bloch equation. The Ba\v{n}as equation is supposed to be used in a continuum setting with sub-grain discretization by the finite-element method. Thus, local geometric features and nonuniform magnetic states during switching are taken into account. We implement the Ba\v{n}as model and test its capability for predicting the recording performance in a realistic recording scenario. By performing recording simulations on 100 media slabs with randomized granular structure and consecutive read back calculation, the write position shift and transition jitter for bit lengths of 10nm, 12nm, and 20nm are calculated.
\end{abstract}

}
\maketitle
\section{Introduction}
Heat assisted magnetic recording (HAMR) is one of the most promising techniques to increase the areal density of magnetic hard disk drives. In perpendicular magnetic recording (PMR) bits are written on a granular medium by magnetizing a cluster of grains in up and down direction. To increase the areal density, the size of the grains and the number of grains per bit have to be reduced. This requires material with higher magnetocrystalline anisotropy to ensure the thermal stability of the written bits. The bits can be written by a moderate writing field selectively by applying a local heat spot, which reduces the coercivity only in the relevant grains. 

Micromagnetic models are an important tool to explore the possibilities and optimal design of new recording technologies. The recording process for perpendicular magnetic recording is often simulated by performing finite-element calculations using the Landau-Lifshitz-Gilbert (LLG) equation. One premise for this model is a constant saturation magnetization $M_\mathrm{s}$ over time. In HAMR the variation of temperature in space and time implies a variation of $M_\mathrm{s}$ and other magnetic properties. Therefore, multiple other approaches were suggested.

The effect of elevated temperature on magnetization dynamics was initially taken into account through an additional random thermal field to the effective field of the LLG equation~\cite{Lyberatos1993,Garcia1998}. Here, each discretization cell was assumed to be one macro spin. Due to the nature of LLG, the norm of the magnetization is conserved. As a consequence of the constant norm and the truncation of high frequency spin waves on a discretization length, which is larger than the atomistic lattice, the Curie temperature $\tau_\mathrm{C}$ is significantly overestimated as pointed out in~\cite{Grinstein2003}. 

Garanin~\cite{Garanin1997} derived the Landau-Lifshitz-Bloch (LLB) equation, which accounts for thermal fluctuation of exchange coupled spins at the atomistic level. The equation of motion includes not only the transverse relaxation of the magnetization vector towards the effective field (as in the LLG), but also the temperature-dependent longitudinal relaxation, which is the magnitude of the magnetization. Both relaxation terms are tuned by phenomenologically found damping parameters. LLB is valid below and above the Curie temperature~\cite{Kazantseva2008}. Therefore, LLB is commonly used in order to study magnetic recording at elevated temperature~\cite{Chubykalo2006,Greaves2012,McDaniel2012, Zhu2013}. As an input of the LLB, temperature dependence of the intrinsic material parameters (saturation magnetization $M_\mathrm{s}$, exchange constant $A_\mathrm{x}$, magnetocrystalline anisotropy constant $K_\mathrm{u}$) and the parallel and perpendicular susceptibility is required. These parameters are usually obtained from simulations using LLG at the atomistic level.

A different approach for the treatment of thermal effects was proposed by Victora and Huang~\cite{Victora2013}. The temperature-dependent material parameters are numerically found on different length scales using atomistic simulations. The length scale-dependent parameters can be used as an input for the material parameters for different mesh-sizes. 

Recently, Ba\v{n}as \etal/~\cite{Banas2009, Banas2012} presented an unconditionally convergent numerical scheme for the approximate solution of LLB. They show that it is possible to eliminate the damping parameter of the longitudinal relaxation term when the temperature dependence of the saturation magnetization is known. Thus, the ambiguity that this additional damping parameter introduced to the simulations is removed.

In this paper, we implement the model proposed by Ba\v{n}as \etal/~\cite{Banas2012} where a LLB type equation is used to incorporate the thermal changes; 100 media slabs with random granular structure closely resembling current media are generated. Recording simulations are performed on these media slabs with three different bit lengths. Based on the reciprocity principle, the read back signals of all recorded tracks are calculated. From these data, we calculate the transition jitter for each bit length. The write position shift for our specific setup is also determined.

\section{Method}
\subsection{Model}
We implemented the equation by Ba\v{n}as \etal/~\cite{Banas2012} who proposed a model, where the LLG equation is extended by an additional term $\kappa \vect{m}$. Therefore, the equation is turned into a LLB type equation, where said term is scaling the magnetization $\vect{M}$ according to the temperature $\tau$
\begin{equation}
\frac{\diff \vect{M}}{\diff t} = \kappa \vect{M} - \frac{\gamma}{1+\alpha^2}\vect{M}\times\vect{H}_\mathrm{eff} - \frac{\gamma\alpha}{1+\alpha^2}\frac{\vect{M}}{M_\mathrm{s}} \times \left(\vect{M}\times\vect{H}_\mathrm{eff}\right)\label{eq:llb}
\end{equation}
where $\gamma$ is the gyromagnetic ratio, $\alpha$ is the damping coefficient, $M_\mathrm{s}$ the saturation magnetization, and $\vect{H}_\mathrm{eff}$ the effective field. The change of the saturation magnetization with temperature can be described by the phenomenological Landau power law 
\begin{equation}
M_\mathrm{s}\left(\tau\right)=M_0\left(1- \frac{\tau}{\tau_\mathrm{C}}\right)^\beta\label{eq:pl}
\end{equation}
with $\tau<\tau_\mathrm{C}$ and $\beta>0$. Here, $M_0$ denotes the saturation magnetization at $\tau=\SI{0}{K}$, $\beta$ is an experimentally found exponent, and $\tau_\mathrm{C}$ is the Curie temperature of the material. The scaling function $\kappa$ is derived from this relation~\cite{Banas2009} yielding
\begin{equation}
\kappa\left(\tau,\tau_\mathrm{C}\right)=-\frac{\diff \tau}{\diff t}\frac{\beta}{\tau_\mathrm{C}-\tau}.\label{eq:kappa}
\end{equation}
The $\kappa$ function, therefore, introduces longitudinal relaxation by incorporating the temperature in space and time and also its time derivative, while satisfying the power law~\eqref{eq:pl}.

Defining $\vect{m}=\vect{M}/M_0$, $\gamma'=\gamma/(1+\alpha^2)$ and $\vect{h}_\mathrm{eff}=\mu_0\vect{H}_\mathrm{eff}$ \eqref{eq:llb} is reformulated into
\begin{equation}
\frac{\mu_0}{\gamma'}\frac{\diff\vect{m}}{\diff t} = \frac{\mu_0}{\gamma'}\kappa\vect{m}-\vect{m}\times\vect{h}_\mathrm{eff}-\alpha\vect{m}\frac{M_0}{M_\mathrm{s}}\times\left(\vect{m}\times\vect{h}_\mathrm{eff}\right).\label{eq:normllb}
\end{equation}
Note that $\vect{m}$ is not the unit vector of the magnetization, but rather describes the scaling of the magnetization with respect to the maximum saturation magnetization. To simplify the equation, we introduce a new timescale as $t^*=t\gamma'/\mu_0$ and define $\kappa^*=\kappa\mu_0/\gamma'$.
Ba\v{n}as \etal/~\cite{Banas2014} suggest to scale the transversal damping constant $\alpha$ in order to obtain an agreement with the stochastic LLG equation. A similar temperature-dependent damping in micromagnetics with a mesh size large than the atomic distances was found by Victora and Huang~\cite{Victora2013}. In this paper, we implement the temperature-dependent transversal damping constant according to numerical experiments reported in~\cite{Banas2014}. We replace $\alpha$ by $\alpha\nu^{-0.75}$, where $\nu$ is the normalized saturation magnetization from \eqref{eq:pl}
\begin{equation}
\nu(\tau) =\frac{M_\mathrm{s}\left(\tau\right)}{M_0}=\left(1-\frac{\tau}{\tau_\mathrm{C}}\right)^\beta.\label{eq:nu}
\end{equation}
With the new timescale, the scaled transversal damping, and the normalized saturation magnetization plugged in, we get the final form of the equation
\begin{equation}
\frac{\diff \vect{m}}{\diff t^*} = \kappa^* \vect{m} - \vect{m}\times\vect{h}_\mathrm{eff} - \frac{\alpha}{\nu^{1.75}} \vect{m}\times \left(\vect{m}\times\vect{h}_\mathrm{eff}\right).\label{eq:banas}
\end{equation}

So far the temperature dependence of $M_\mathrm{s}$ and $\alpha$ was taken care of. Other temperature-dependent intrinsic properties, exchange coupling constant $A_\mathrm{x}(\tau)$, and uniaxial anisotropy constant $K_\mathrm{u}(\tau)$ are included in the respective field terms of the effective field $\vect{h}_\mathrm{eff}$: the exchange field $\vect{h}_\mathrm{xhg}$ and the anisotropy field $\vect{h}_\mathrm{ani}$. They are given by
\begin{equation}
 \vect{h}_\mathrm{xhg}=\frac{2A_\mathrm{x}}{M_\mathrm{s}^2}\nabla^2\vect{M}\quad\text{ and } \quad
 \vect{h}_\mathrm{ani}=\frac{2K_\mathrm{u}}{M_\mathrm{s}}\vect{k}\left(\vect{u}\vect{k}\right)\label{eq:term}
\end{equation}
with $\vect{k}$ being the unit vector of the magnetocrystalline anisotropy easy axis and $\vect{u}=\vect{M}/M_\mathrm{s}$ the unit vector of the magnetization $\vect{M}$. 
Different temperature dependencies of $A_\mathrm{x}$, $K_\mathrm{u}$, and $M_\mathrm{s}$ were proposed usually being proportional to $\left(M_\mathrm{s}\left(\tau\right) / M_0\right)^\eta=\nu^\eta\left(\tau\right)$ with different values for the exponent $\eta$~\cite{Richter2012,Bublat2010,Greaves2012a}. For simplicity, in this paper we set $\eta=2$. With this choice, we approximately have $K_\mathrm{u}(\tau)\sim M(\tau)^2$, which holds for FePt films, as shown in~\cite{Mryasov2005}. In our model, the exchange constant changes with temperature as $A_\mathrm{x}(\tau)\sim M(\tau)^2$. The exponent corresponds to the mean field value and is higher than the value of 1.76 obtained for FePt at low temperature~\cite{Atxitia2010}. Therefore, with $A_\mathrm{x}(\tau)=A_0\nu^2(\tau)$ and $K_\mathrm{u}(\tau)=K_0\nu^2(\tau)$ with $A_0$ and $K_0$ being the properties at $\tau=\SI{0}{K}$, both field terms in \eqref{eq:term} can be reformulated to
\begin{equation}
 \vect{h}_\mathrm{xhg}=\frac{2A_0}{M_0}\nabla^2\vect{m}\quad\text{ and } \quad
 \vect{h}_\mathrm{ani}=\frac{2K_0}{M_0}\vect{k}\left(\vect{m}\vect{k}\right).
\end{equation}
One can see that $\nu(\tau)$ vanishes and the dependence on $\tau$ is conveniently introduced by the already scaled $\vect{m}$ in both field terms. The demagnetization field is not directly dependent on $\tau$ and hence can be calculated as usual with the magnetic potential $\nabla^2\phi=\nabla \vect{M}$ by $\vect{h}_\mathrm{dmg}=-\nabla\phi$. Finally, with the externally applied field $\vect{h}_\mathrm{ext}$, the effective field is given by
\begin{equation}
\vect{h}_\mathrm{eff}=\vect{h}_\mathrm{xhg}+\vect{h}_\mathrm{ani}+
\vect{h}_\mathrm{dmg}+\vect{h}_\mathrm{ext}.\label{eq:heff}
\end{equation}

\subsection{Implementation}
We developed a finite element c++ code and implemented the proposed scheme. By using a sparse linear algebra library for OpenCL~\cite{Demidov2013}, the recording simulations can be performed massively parallel on graphics processors to reduce computation time. Time discretization is done by a mid-point rule with a fixed time step of \SI{5}{fs}. In each time step, the resulting nonlinear equation is solved by a quasi-Newton method~\cite{Li2007} using a limited memory Broyden-Fletcher-Goldfarb-Shanno (L-BFGS) update~\cite{Liu1989} and a derivative-free line search with the Griewank condition~\cite{Griewank1986}. 

\begin{figure}[htb]
\centering
\includegraphics[width=\linewidth]{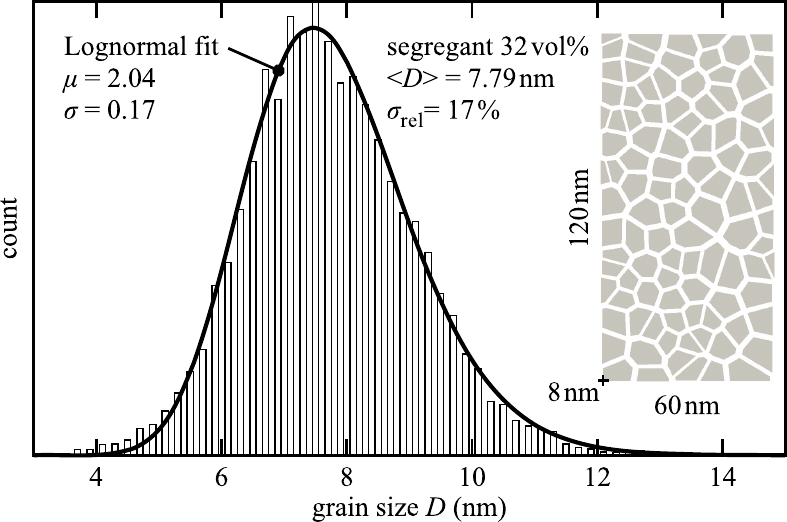}
\caption{Lognormal fit (solid line) of the used grain size distribution of 100 media slabs with dimensions \SI{120x60x8}{\nano\metre}. Each slab consist of 100 randomly generated grains, one example is shown on the right hand side.}
\label{fig:01}
\end{figure}

Models of magnetic storage media are generated with random granular structure. By using the polycrystal generation tool Neper~\cite{Quey2011}, the morphology can be tuned to resemble the grain size distribution of current media for HAMR~\cite{Weller2014,Weller2016}. We created 100 different realizations of a \SI{120x60x8}{\nano\metre} media slab, each consisting of 100 grains with a mean grain size of $\left\langle D\right\rangle=\SI{7.8}{nm}$ (cf. Fig.~\ref{fig:01}) to perform recording simulations. A nonmagnetic intergranular phase separates the magnetic grains by \SIrange{0.5}{2.5}{nm} and takes about \SI{32}{vol\%} of the medium. The medium is embedded in a \SI{800x400x72}{\nano\metre} air box for stray field computation. Both, the intergranular phase and the surrounding air box are generated within Salom\'{e}~\cite{salome}, a computer aided design software. The whole model is discretized by a tetrahedral finite element mesh, which is generated with Netgen~\cite{Schoberl1997}. The mesh size for the magnetic grains is set to \SI{4}{nm} with a growing mesh size for the surrounding air box. 

The material properties for the $\mathrm{L}1_0$ FePt grains were set to $J_\mathrm{s}=\SI{1.43}{T}$ saturation polarization, $\langle K_\mathrm{u}\rangle=\SI{6.6}{MJ/m^3}$ uniaxial magnetocrystalline anisotropy constant with just \SI{1}{\%} standard deviation, and $A_\mathrm{x}=\SI{2.158}{pJ/m}$ exchange stiffness constant at $\tau=\SI{300}{K}$. $A_\mathrm{x}$ is taken from~\cite{Vogler2016} and is valid within the grains, the inter-granular phase is nonmagnetic, and therefore the grains are coupled only magnetostatically. The Curie temperature is assumed to be $\tau_\mathrm{C}=\SI{830}{K}$, and the exponent for the temperature dependence of $M_\mathrm{s}$ is set to $\beta=0.36$. The transversal damping constant at \si{300}{K} is set to $\alpha=0.1$. The out-of-plane easy axes of the grains are well aligned.

\begin{figure}[htb]
\centering
\includegraphics[width=\linewidth]{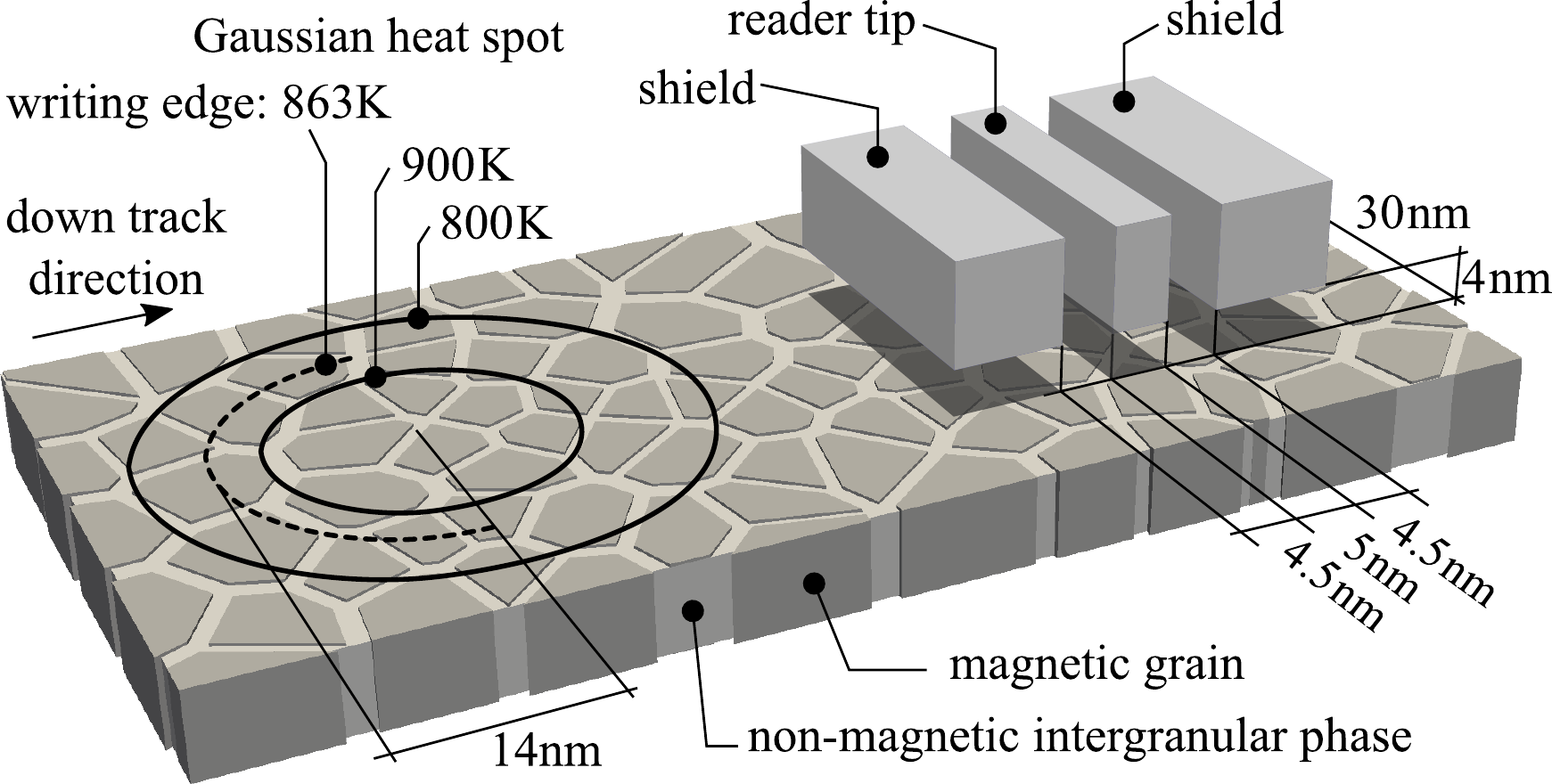}
\caption{Schematic of a media slab showing the relative size of the heat spot on the left with FWHM=\SI{75}{nm} and a peak temperature of $\tau_\mathrm{max}=\SI{950}{K}$ used to record the bits. The dashed half circle marks the writing edge. The layout of the read head is shown on the right-hand side.}
\label{fig:02}
\end{figure}

Experimental data of a heat spot were fitted by a Gaussian 2-D surface with $\sigma_x=\sigma_y=\SI{32}{nm}$ and a peak temperature of $\tau_\mathrm{max}=\SI{950}{K}$. The size of the heat spot relative to the media slab is shown at the left-hand side of Fig.~\ref{fig:02}. 
The bit sequences are recorded on initially completely upward magnetized media slabs. During the recording simulation, the continuous heat spot moves with a velocity of \SI{20}{m/s} across the medium model in the down-track direction. The ambient temperature is set to \SI{300}{K}. A global homogeneous field of \SI{1}{T} at an angle of \SI{45}{\degree} is applied. By alternating the fields polarity with both the rise and decay time being \SI{0.1}{ns}, an alternating bit sequence is written on each medium. The pulse width is set according to the desired bit length (cf. recording field as dashed-dotted line in Fig.~\ref{fig:03}).

The read back signal is computed by a reciprocity reader. The thickness of the reader tip is \SI{5}{nm}, the gap between shield and tip \SI{4.5}{nm} (shield to shield spacing of \SI{14}{nm}), and the flight height is \SI{4}{nm} above the medium. A schematic of the reader is shown at the right-hand side of Fig.~\ref{fig:02}. For the read back calculation, only a track width of \SI{30}{nm} along the center axis in down-track direction of the media is considered. In Fig.~\ref{fig:03}, this area is marked in the sample track insets by the dashed lines.
The transition jitter $\sigma_\mathrm{trans}$ is the standard deviation of written bit length from the desired bit length $l_\mathrm{b}$. It is calculated by using the $N$ distances between neighboring zero transitions $\delta_i=\zeta_i-\zeta_{i-1}$ with $i=1, \dots, N$ of the read back signals. For $N+1$ transitions, we get
\begin{align}
\sigma_\mathrm{trans}&=\sqrt{\frac{1}{N-1}\sum_{i=1}^N{\left(\delta_i-l_\mathrm{b}\right)^2}}.\label{eq:jitter}
\end{align}

\section{Results}
By using the previously described Ba\v{n}as model, we performed recording simulations for three different bit lengths $l_\mathrm{b}$ of \SIlist{10; 12;20}{\nm}. For each bit length, the bit transitions where recorded on 100 media slabs with randomly varying granular structure. The pulse width of the external field was \SIlist{0.3;0.4;0.8}{\ns} for the bit lengths \SIlist{10;12;20}{\nm} respectively. 
\begin{figure}[htb]
\centering
\includegraphics[width=\linewidth]{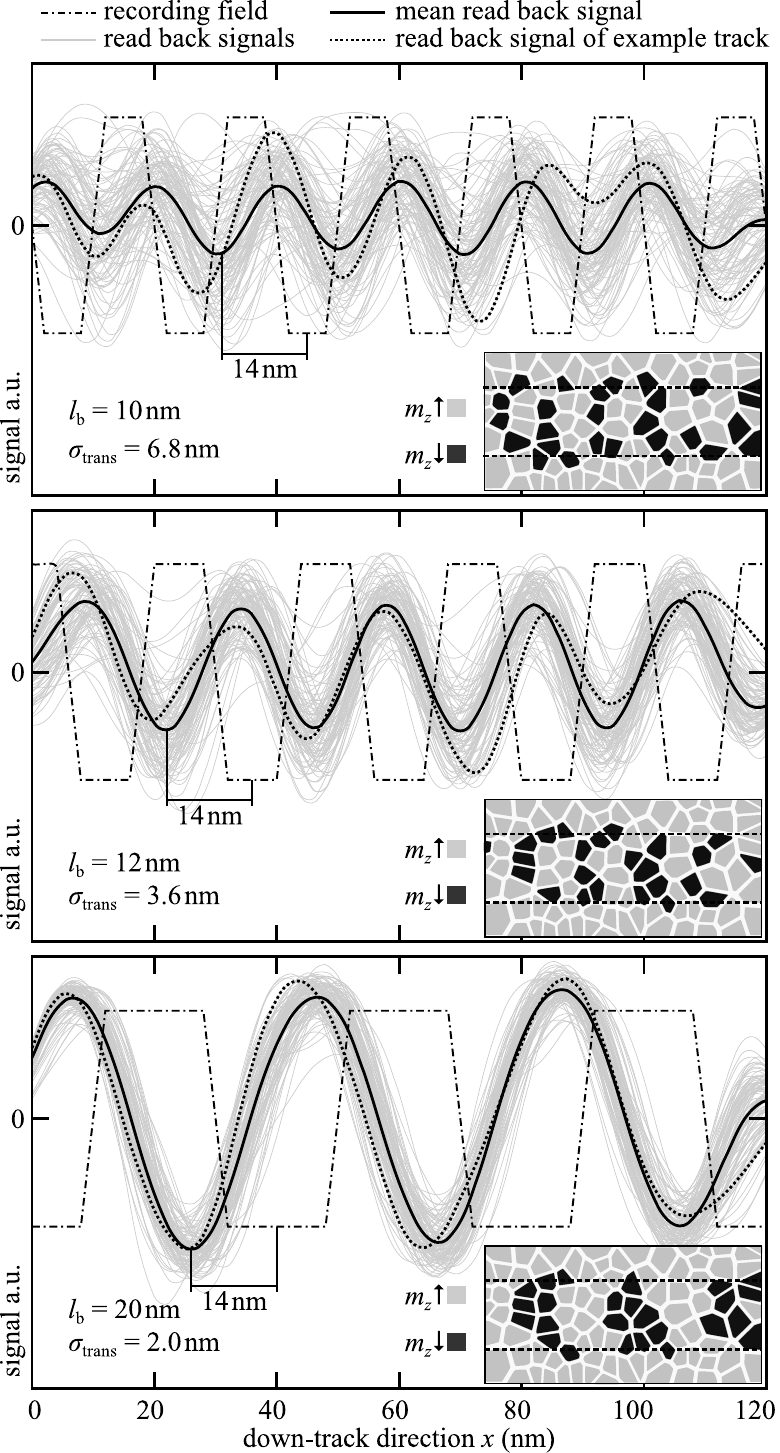}
\caption{The diagrams show the mean read back signal as a black solid line of 100 recording simulations with varying random granular structure for bit length \SI{10}{nm} (top), \SI{12}{nm} (center), and \SI{20}{nm} (bottom). The recording field is shown with respect to the heat spot maximum as a black dashed-dotted line and the various read back signals in thin gray lines. One example of a recorded track is shown in each bottom right corner with the respective signal as dotted line. The reader only considers the \SI{30}{nm} wide track between the dashed lines drawn on the media. The writing edge of the heat spot is \SI{14}{nm} behind the heat spot maximum.}
\label{fig:03}
\end{figure}

For each bit length the various read back waveforms for the different media realization are shown as ensemble of gray lines in Fig.~\ref{fig:03}. The mean of the 100 signals is shown as a solid black line. The phase shift between the recording field signal (dashed-dotted line) and the read back signal appears because the field signal is shown with respect to the $x$-position of the heat spot peak. The bits are written with the back edge of the heat spot and the external field provides only the magnetization direction. Therefore, the phase shift is actually the distance of the writing edge to the peak of the heat spot. In the presented simulations, the shift is \SI{14}{nm}, which translates to a writing edge of \SI{863}{K} (cf. Fig.~\ref{fig:02}).

One of the 100 recorded tracks for each bit length is shown as inset in each diagram. The corresponding signal is highlighted as dotted line. While at $l_\mathrm{b}=\SI{20}{nm}$, one bit is stored on roughly nine grains, at $l_\mathrm{b}=\SI{10}{nm}$, one bit occupies about five grains. This leads of course to a lower read back signal with decreasing bit length. Moreover, with decreasing bit length, some bit transitions are not picked up by the reader head as a zero transition. This can be seen in Fig.~\ref{fig:03} (top) 	for $l_\mathrm{b}=\SI{10}{nm}$ in the signal of the shown track (dotted line) between \SIlist{80;100}{nm}. This, of course, increases the jitter value. In order to calculate jitter, we cut away the first and last written bit of the signal of all bit lengths to avoid distorting effects on the media slabs edges. From these data, the position jitter was calculated to be \SIlist{6.8;3.6;2.0}{\nm}, for a bit length of $l_\mathrm{b} = \SI{10}{nm}, \SI{12}{nm}$, and $\SI{20}{nm}$, respectively.

\section{Conclusion}
Our implementation of the Ba\v{n}as equation can be used to simulate the heat-assisted recording process for temperatures below the Curie temperature. 
The calculated jitter values for the three investigated bit lengths compare well to those from literature~\cite{Victora2013,Wu2013}. 
Besides the transition jitter, also the position of the writing edge is regarded as a characteristic parameter for HAMR. Changing the writing head velocity, the heat spot design or the head field will influence this write position shift~\cite{Wang2017}. With the here presented model, this parameter can be determined and optimized when designing a HAMR system.

\section*{Acknowledgment}
This work was supported by the Vienna Science and Technology Fund (WWTF) under Grant MA14-044.

\IEEEtriggeratref{25}
\bibliographystyle{IEEEtran}
\bibliography{ref}
\end{document}